\documentclass[aps,prx,twocolumn,floatfix,footinbib,showpacs,superscriptaddress]{revtex4-1}
\usepackage{graphicx}
\usepackage{amsfonts,amsmath,amssymb}
\usepackage{amsthm}
\usepackage{dsfont}
\usepackage{bm}%
\usepackage{color}
\usepackage{soul} %Include when comparing comments
\usepackage{amsbsy}
\usepackage{upgreek}

\usepackage[colorlinks=true,linkcolor=blue,pagecolor=blue,filecolor=blue,menucolor=blue,urlcolor=blue,citecolor=blue,anchorcolor=blue]{hyperref}%

\newcommand{\bsigma}{{\boldsymbol\upsigma}}

\begin{document}

%\preprint{APS/123-QED}

\title{Half-Metallic Superconducting Triplet Spin Valve}
%\title{Spin Controlled Triplet Correlations in a Half-Metallic Spin Valve}
\author{Klaus Halterman} \email{klaus.halterman@navy.mil}
\affiliation{Michelson Lab, Physics Division, Naval Air Warfare Center, China Lake, California 93555}
\author{Mohammad Alidoust} \email{phymalidoust@gmail.com}
%\affiliation{Department of Physics, University of Basel, Klingelbergstrasse 82, CH-4056 Basel, Switzerland}
\affiliation{Department of Physics, Faculty of Sciences, University of Isfahan, Hezar Jerib Avenue, Isfahan 81746-73441, Iran}
\date{\today}

\begin{abstract}
We theoretically study a finite 
size $SF_1NF_2$ spin valve, 
where a normal metal ($N$) insert separates 
a  thin  standard ferromagnet ($F_1$)  and a thick half-metallic ferromagnet  ($F_2$).
For sufficiently thin superconductor ($S$)  widths close to the coherence length $\xi_0$, we find
that changes to the relative magnetization orientations in the ferromagnets can result in
substantial variations
in the transition temperature $T_c$,
consistent with experiment [Singh {\it et al.,} \href{http://journals.aps.org/prx/abstract/10.1103/PhysRevX.5.021019}{Phys. Rev. X {\bf 5},
021019 (2015)}]. Our results demonstrate that, in good agreement with
the experiment, the variations are
largest in the case where $F_2$ is in a half-metallic phase and thus
supports only one spin direction. 
To pinpoint the origins of
 this strong spin-valve effect,
 both the equal-spin $f_1$  and opposite-spin $f_0$  triplet 
correlations are calculated 
 using a self-consistent microscopic technique. 
 We find that when the magnetization
 in $F_1$ is tilted slightly out-of-plane,
 the
 $f_1$  component can be the dominant triplet component 
in the superconductor. The coupling between the two ferromagnets is 
discussed in terms of the 
underlying spin currents present in the system.
We go further and show that the zero energy peaks of the local density of states
probed on the $S$ side of the valve can be another signature of the
presence of
superconducting triplet correlations.
Our findings reveal that for sufficiently thin $S$ layers,
the zero energy peak at the $S$ side can be larger
than its counterpart in the $F_2$ side. 
\end{abstract}


\maketitle

In the field of superconducting spintronics, 
there is interest in spin-controlled proximity effects 
 for  manipulating the superconductivity  in 
ferromagnet ($F$) and superconductor ($S$) layered
systems \cite{review,Eschrig2015}. 
When an $S$ layer is
 in contact with two ferromagnets, 
creating a  superconducting spin valve, 
the superconducting state can be controlled by changing 
the relative magnetization directions \cite{fomin,karmin1,wu}.
The  basic superconducting spin-valve 
 involves
 $SFF$ structures \cite{lek,fomin}
where 
switching between relative 
parallel and antiparallel magnetizations  modifies
 the oscillatory
 singlet pairing in the $F$ regions.
For strong ferromagnets, these oscillations have  limited extent,
as they become damped out
over very short distances \cite{Eschrig2008}.
If however,
the 
mutual magnetizations vary  noncollinearly,
the broken time reversal and translation
symmetries
 induces a mixture of
spin singlet
and odd-frequency (or odd-time) 
 spin-triplet
correlations with $0$ and $\pm 1$ spin projections along the
magnetization axis \cite{first,Buzdin2005}. 
The 
triplet pairs with nonzero 
spin projection can
 naturally penetrate extensively
within  the ferromagnet layers \cite{Keizer2006,Halterman2007, Halterman2008, Bobkova,Moor,Khaydukov,longrg} 
%ref1_p3 ->
and result in an enhancement of  the DOS 
at low energies \cite{buz_zep,berg_zep}.
%ref1_p3 <-
This long-range triplet component in $S F_1 F_2$ type spin valves
 can be manipulated by changing the relative orientations
of the magnetizations in $F_1$ and in $F_2$, which creates opportunities for
the development of new types of  spin-valves and switches for
nonvolatile memory applications \cite{karmin2,Bakurskiy1,Bakurskiy2}.
Because of their simplicity in pinpointing fundamental phenomena and
promising prospects in spintronics devices, the $S F_1 F_2$ spin
valve continues to attract broad interest \cite{karmin2,singh,fomin,nowak,lek,bernard,golu,longrg,klaus_zep,Mironov}. 
For example, an
anomalous Meissner effect
has recently been observed \cite{meissner_exp} that is consistent with
the generation of an odd-frequency superconducting state \cite{mkj}.

Recent experiments involving  superconducting
spin valves have 
investigated variations in the critical temperature, $T_c$ \cite{ilya,jara} 
when varying the relative
in-plane magnetization angle.
The suppression in $T_c$
for nearly orthogonal 
magnetizations reflects
the increased presence of equal-spin triplet pairs \cite{lek}.
A spin valve like effect
was also experimentally realized \cite{west,nowak}  in FeV superlattices,
where antiferromagnetic coupling between the Fe layers permits
gradual rotation of the relative magnetization direction in
the $F_1$ and $F_2$ layers.
Most experiments involve standard ferromagnets,
leading to $\Delta T_c$ sensitivity of several mK. 
When the outer $F_2$ layer is replaced by a  half-metallic ferromagnet, such as
${\rm Cr O}_2$,
a very large  $\Delta T_c$  has been reported, which is indicative of
the presence of odd-frequency triplet superconducting
correlations \cite{singh}.

\begin{figure}
\centering
\includegraphics[width=0.9\columnwidth]{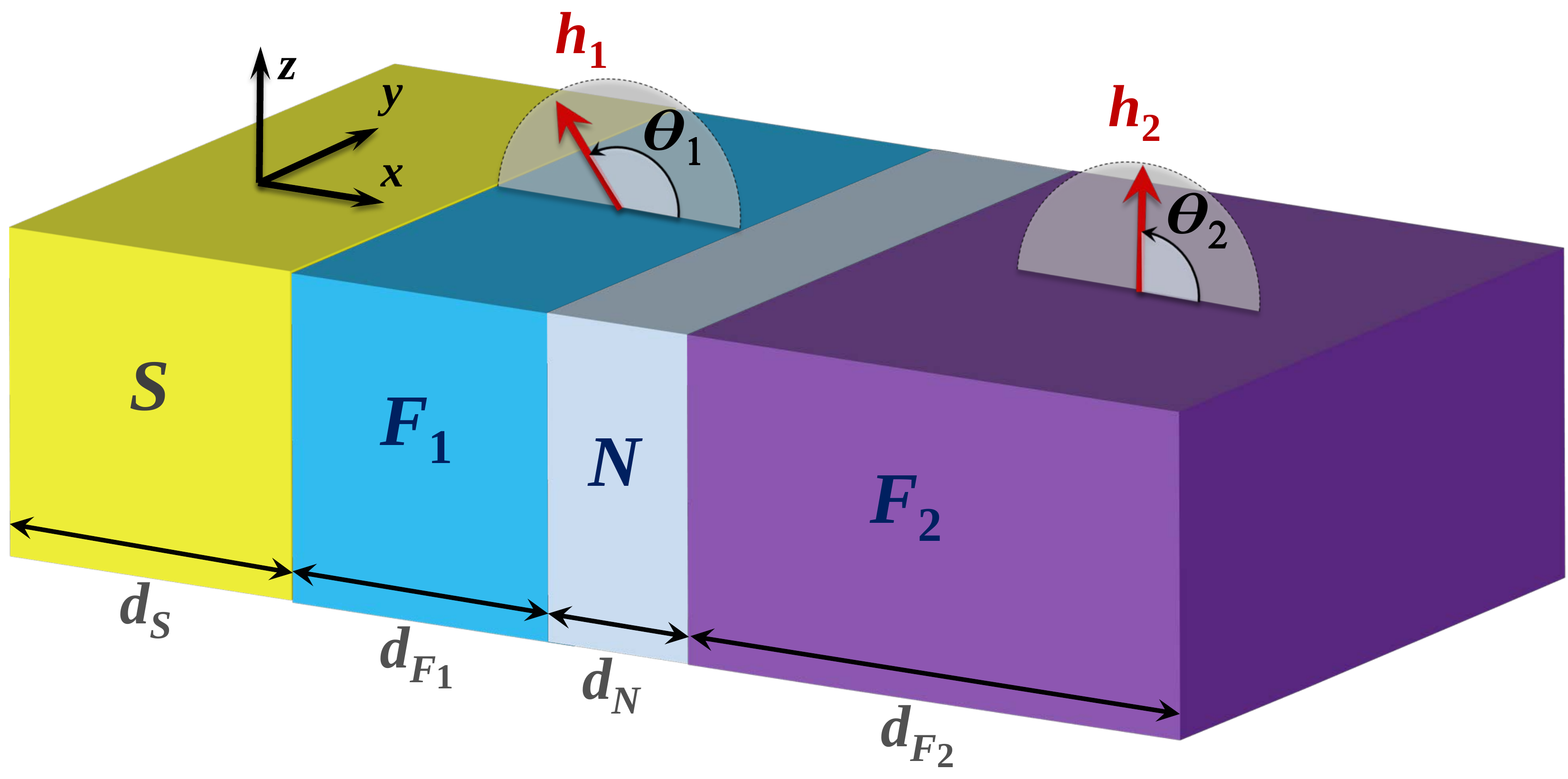}
\caption{(Color online). Schematic of the finite size $SF_1NF_2$ multilayer, 
where $\theta_1$ and $\theta_2$
characterize  the magnetization orientation of ferromagnets $F_1$ and
$F_2$ with thicknesses $d_{F_1}$ and $d_{F_2}$, respectively.
The normal metal ($N$) insert with thickness $d_N$ is a nonmagnetic layer such as ${\rm Cu}$.
The exchange field in each magnet is written ${\bm h}_i=h_i(\cos\theta_i,0,\sin\theta_i)$,
for $i=1,2$. Here, $\theta_i$ is measured relative to the $x$-axis.
The ferromagnet $F_2$ is half-metallic (e.g., ${\rm Cr O}_2$) so that $|{\bm h}_2| = E_F$, and its magnetization
is fixed along the $z$ direction ($\theta_2=\pi/2$), whereas the magnetization in $F_1$
can rotate in the $x-z$ plane.
We thus define the  angle $\theta$
to describe the out-of-plane relative magnetization between the two magnets,
with $\theta\equiv \theta_1-\theta_2$.
}
\label{schematic}
\end{figure}

Besides through studying $T_c$,
the existence and type of superconducting
correlations in superconducting spin-valves can be identified through 
signatures of the proximity-induced electronic density
of states (DOS) \cite{naz}.
When  triplet 
correlations are present in an $F$ layer, 
it has been shown that
a zero energy peak (ZEP) in the DOS can arise \cite{golubov,klaus_zep}.
The situation where pair correlations 
 from both the spin-0 and spin-1 triplet channels
 are present
can  however make its unambiguous detection
difficult.
Nonetheless, this 
difficulty can be alleviated if one of the $F$ layers is
half-metallic (supporting one spin direction), creating an effective spin-filter
that can isolate 
 the spin-1 triplet component due to
the large exchange splitting present.
Thus it  is of interest to investigate  $S F_1 F_2$ structures
containing a half-metallic ferromagnet,
where the modified triplet  proximity effects can
result in strong spin valves with high sensitivity to magnetization changes
and a corresponding $T_c$ suppression. 

%khx1 ------
To realistically and accurately  model
 these systems, where $h\simeq E_F$, we 
 use
a fully  microscopic microscopic framework, 
the  Bogoliubov-de Gennes (BdG) equations,
to  determine the singlet and triplet pair correlations self-consistently.
%ref1_p1 -> %khf
This approach   naturally supports 
the study  of a broad range of intermediate ferromagnetic
exchange energies, including
the half metallic phase, 
by simply setting the exchange field 
value close to the Fermi energy.
The half metallic regime is  also
accessible within the quasiclassical approximation \cite{eschrig_half,Mironov}
 by considering the case when the energy splitting of the the spin-up
 and spin-down bands greatly exceed the Fermi energy, i.e., $h\gg E_F$.
Using the BdG formalism,
we  show how to identify the existence of the equal-spin triplet
components  by probing the $S$ side of the proposed valve with an STM, revealing 
signatures in the form of peaks in the  density of states (DOS) at zero
energy \cite{klaus_zep,bernard}.
 
\section{methods}
A schematic
of the spin valve configuration is depicted in Fig.~\ref{schematic}. 
 We model the  
nanostructure
as  a
${SF_1NF_2}$ layered system, 
where $S$ represents the superconducting layer,
$N$ denotes the normal metallic intermediate layer,
and $F_1$, $F_2$ 
are the inner (free) and outer (pinned)
magnets, respectively.  
The 
layers are assumed to be infinite in the $y-z$ plane with a total thickness 
$d$ in the $x$ direction, which is perpendicular to the interfaces between layers. 
The ferromagnet 
$F_2$ has width $d_{F_2}$, and
fixed direction of magnetization along $z$,
while
the free magnetic layer $F_1$ of width $d_{F_1}$ 
has a variable magnetization
direction. 
The superconducting layer of thickness $d_S$ is in contact with the free layer. 
The  magnetizations in the $F$ layers are modeled by effective Stoner-type exchange fields ${\bm h}(x)$
which vanish in the non-ferromagnetic layers. 

To accurately describe the physical properties of our systems with sizes in the 
nanometer scale and over a broad range of exchange fields, 
where quasiclassical approximations 
are limited,  
we numerically 
solve the microscopic BdG equations within a fully self-consistent framework. 
The general
spin-dependent BdG equations for the quasiparticle energies, $\varepsilon_n$, 
and quasiparticle wavefunctions, $u_{n\sigma}, v_{n\sigma}$ is written:
\begin{align}
&\begin{pmatrix} 
{\cal H}_0 -h_z&-h_x&0&\Delta(x) \\
-h_x&{\cal H}_0 +h_z&-\Delta(x)&0 \\
0&-\Delta(x)&-({\cal H}_0 -h_z)&-h_x\\
\Delta(x)&0&-h_x&-({\cal H}_0+h_z) \\
\end{pmatrix} \nonumber \\
&\times
%H_0\tau_z+{\bm h]\cdot {\bm \sigma}\tau_0+i\Delta\tau_y
\begin{pmatrix} 
u_{n\uparrow}\\u_{n\downarrow}\\v_{n\uparrow}\\v_{n\downarrow}
\end{pmatrix} 
=\varepsilon_n
\begin{pmatrix}
u_{n\uparrow}\\u_{n\downarrow}\\v_{n\uparrow}\\v_{n\downarrow}
\end{pmatrix}\label{bogo2},
\end{align}
where $h_i$ 
($i=x,z$) are components of the 
exchange field. In 
Eqs.~(\ref{bogo2}), 
the single-particle Hamiltonian ${\cal H}_0=
-{1}/{(2m)}{d^2}/{dx^2}-E_F+U(x)$  
contains the Fermi energy, $E_F$, and
an effective interfacial scattering potential described by 
delta functions of strength $H_j$ ($j$ denotes the different interfaces), namely:
$U(x)= H_{1}\delta(x-d_S)$
$+$
$H_{2}\delta(x-d_S-d_{F_1})$
$ +$
$H_{3}\delta(x-d_S-d_{F_1}-d_N)$,
where $H_j={k_F H_{B j}}/{m}$ is written in terms of the dimensionless 
scattering strength $H_{B j}$. 
We assume $h_{x,i}=h_i\cos \theta_i$ and 
$h_{z,i}=h_i\sin \theta_i$  
in $F_i$, where $h_i$ is the magnitude of exchange field, and $i$ denotes the region.  
To minimize the free energy of the system at temperature $T$,
the singlet pair potential $\Delta(x)$ is calculated self-consistently \cite{gennes}:
\begin{equation}   
\label{del} 
\Delta(x) = \frac{g(x)}{2}{\sum_n}
\bigl[u_{n\uparrow}(x)v_{n\downarrow}(x) +
u_{n\downarrow}(x)v_{n\uparrow}(x)\bigr]\tanh\left(\frac{\varepsilon_n}{2T}\right), 
\end{equation} 
where the sum is over all  eigenstates
with $\varepsilon_n$ that lie within a characteristic 
Debye energy $\omega_D$, and  
$g(x)$ is the superconducting coupling strength, taken to be constant in the $\rm S$ region and zero elsewhere. 
The pair potential gives direct  information regarding %khf
superconducting correlations within the $S$ region only,
since it vanishes in the remaining spin valve regions where $g(x) = 0$. 
Greater insight
into the singlet superconducting correlations throughout the structure, and
the extraction of the proximity effects 
is most easily obtained by considering the pair
amplitude, $f_3$, defined as $f_3\equiv \Delta(x)/g(x)$.

To analyze the correlation between the behavior of the 
superconducting transition temperatures and the existence of 
odd triplet superconducting correlations in our system, 
we compute the induced triplet pairing amplitudes which we denote as $f_0$ 
(with $m=0$ spin projection) and $f_1$ (with $m=\pm 1$ spin projection)
according to
the following equations \cite{Halterman2007}:
\begin{subequations}
\begin{eqnarray}
f_0 (x,t) & = \frac{1}{2} \sum_n \left[ u_{n\uparrow} (x) v_{n\downarrow}(x)-
u_{n\downarrow}(x) v_{n\uparrow} (x) \right] \zeta_n(t),\;\;\;\;  \;\;\;
\label{f0defa} \\
f_1 (x,t) & = -\frac{1}{2} \sum_n \left[ u_{n\uparrow} (x) v_{n\uparrow}(x)+ 
u_{n\downarrow}(x) v_{n\downarrow} (x) \right] \zeta_n(t),\;\;\;\; \;\;\;
\label{f1defa}
\end{eqnarray}
\end{subequations}
where $\zeta_n(t) \equiv \cos(\varepsilon_n t)-i \sin(\varepsilon_n t) \tanh(\varepsilon_n /(2T))$,
and $t$ is the time difference in the Heisenberg picture.
These triplet pair amplitudes are odd in $t$ and vanish at $t=0$,
in accordance with the Pauli exclusion principle.
The quantization  axis 
in Eqs.~(\ref{f0defa}) and (\ref{f1defa})
is along the $z$ direction. 
When studying the triplet correlations in $F_1$,
we 
align the quantization axis  
 with the
local
exchange field direction, so that after
rotating, the triplet amplitudes $f_0$ and $f_1$ 
become linear combinations of the $f_0$ and $f_1$
in the original unprimed system \cite{klaus_zep}:
$f_0^\prime(x,t)$$=$ $f_0(x,t) \cos\theta - f_1(x,t) \sin\theta$,
and $f_1^\prime(x,t)$$=$$f_0(x,t) \sin\theta  + f_1(x,t) \cos\theta$.
Thus, when the exchange fields in $F_1$ and $F_2$ are
orthogonal  ($\theta=\pi/2$),
the roles of the equal-spin and opposite-spin triplet correlations are reversed. 
The singlet pair amplitude however is naturally invariant under these rotations.

The study of single-particle 
excitations in these systems can 
reveal important 
signatures  
in the proximity induced singlet and triplet pair correlations.
A useful experimental tool that probes these single-particle states is tunneling
spectroscopy, where information 
measured by a scanning tunneling microscope (STM)
can reveal the local DOS, $N(x,\varepsilon)$, as a function of position $x$
and energy $\varepsilon$. 
We write $N(x,\varepsilon)$ as a
sum of each spin component ($\sigma=\uparrow,\downarrow$)
 to the DOS:
$N(x, \varepsilon) = N_\uparrow(x, \varepsilon) + N_\downarrow(x, \varepsilon)$, 
where,
\begin{eqnarray}
N_\sigma(x, \varepsilon) =\sum_n\left[u_{n\sigma}^2(x) \delta(\varepsilon-\varepsilon_n)+
v_{n\sigma}^2(x) \delta(\varepsilon+\varepsilon_n)
\right].\;\;\;\;\;
\end{eqnarray}

\section{results}
We now proceed to present the self-consistent
numerical  results 
for the transition temperature, triplet amplitudes,
and local DOS for the spin-valve structure depicted in Fig.~\ref{schematic}.
We normalize
the temperature in the calculations  by 
$T_{0}$, the transition
temperature of a pure bulk S sample.
When in the low-$T$ limit, 
we take $T = 0.05 T_{0}$.
All length scales
are normalized by the 
Fermi wavevector $k_F$, 
so that the coordinate
$x$ is written $X = k_F x$, 
and the $F_1$ 
and $F_2$ 
widths are
written 
$D_{F_i} = k_F d_{F_i}$, for $i = 1, 2$.  
The   thick
half-metallic ferromagnet  $F_2$  
has width
$D_{F_2} = 400$, and $F_1$ is a standard ferromagnet 
with $h_1=0.1E_F$.
We set 
 $d_{F_1}=\xi_F$, where $\xi_F=v_F/(2 h_1)$
is the length scale describing  the
propagation of  spin-0 pairs.
 In dimensionless units we thus have,
 $D_{F_1}=(h_1/E_F)^{-1}=10$, 
which optimizes 
spin mixing of superconducting correlations in the system.
The $S$ width is normalized similarly 
by 
$D_{S} = k_F d_{S}$,
and its  scaled  coherence length 
is taken to be $k_F \xi_0 = 100$. 
Natural units, e.g., $\hbar = k_B = 1$, are used throughout.
\begin{figure}
\centering
\includegraphics[width=0.99\columnwidth]{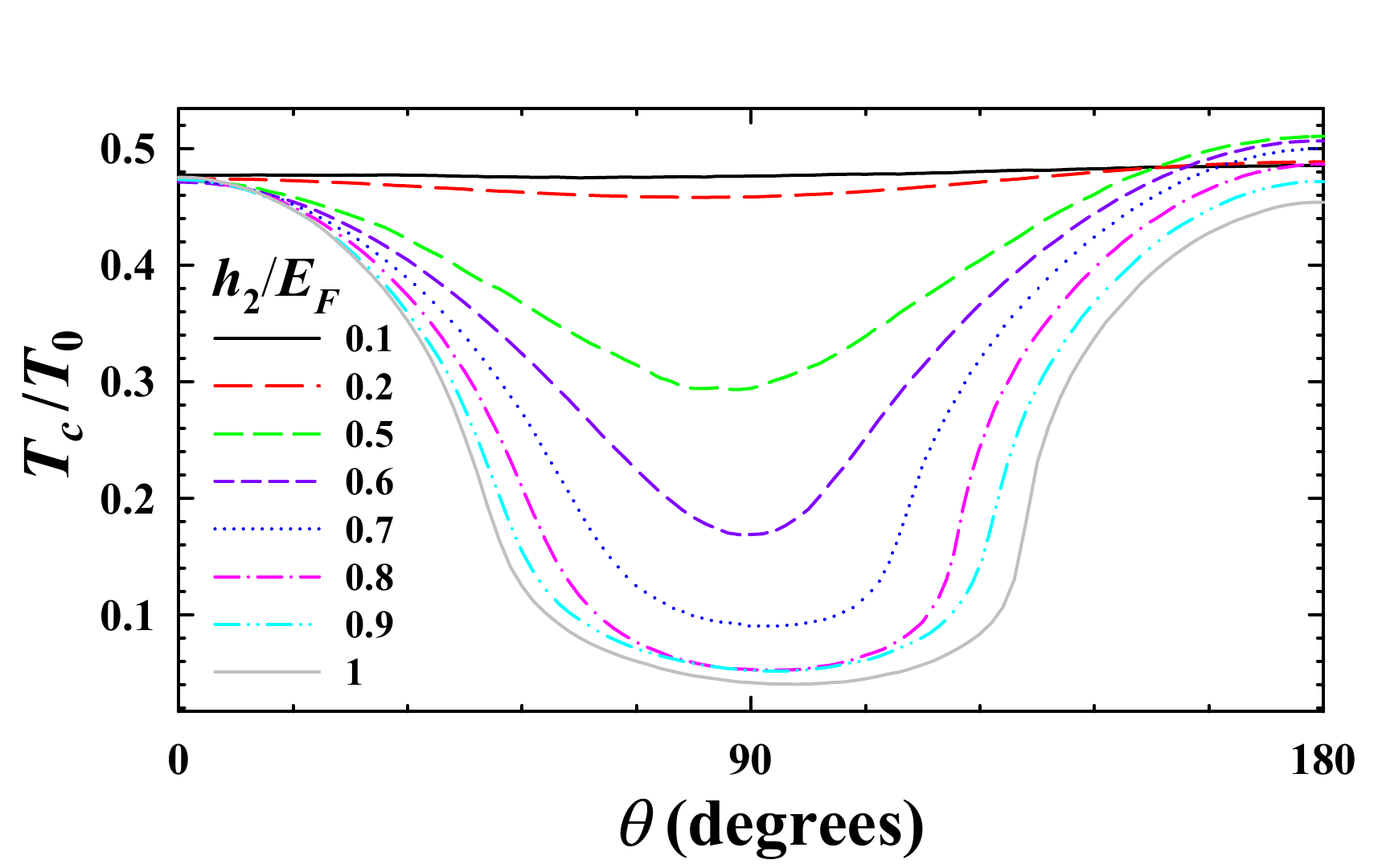}
\caption{
(Color online). Critical temperature $T_c$ as a function of the
relative exchange field orientation angle $\theta$ at differing values
of the 
ratio of the exchange field in the $F_2$ region, $h_2$
to the Fermi energy $E_F$. 
The legend depicts the range of $h_2/E_F$ considered, ranging from 
a relatively weak ferromagnet with $h_2/E_F=0.1$, to 
a fully spin polarized half-metallic phase, corresponding to $h_2/E_F=1$.
}
\label{tc_h}
\end{figure}

\begin{figure}
\centering
\includegraphics[width=0.99\columnwidth]{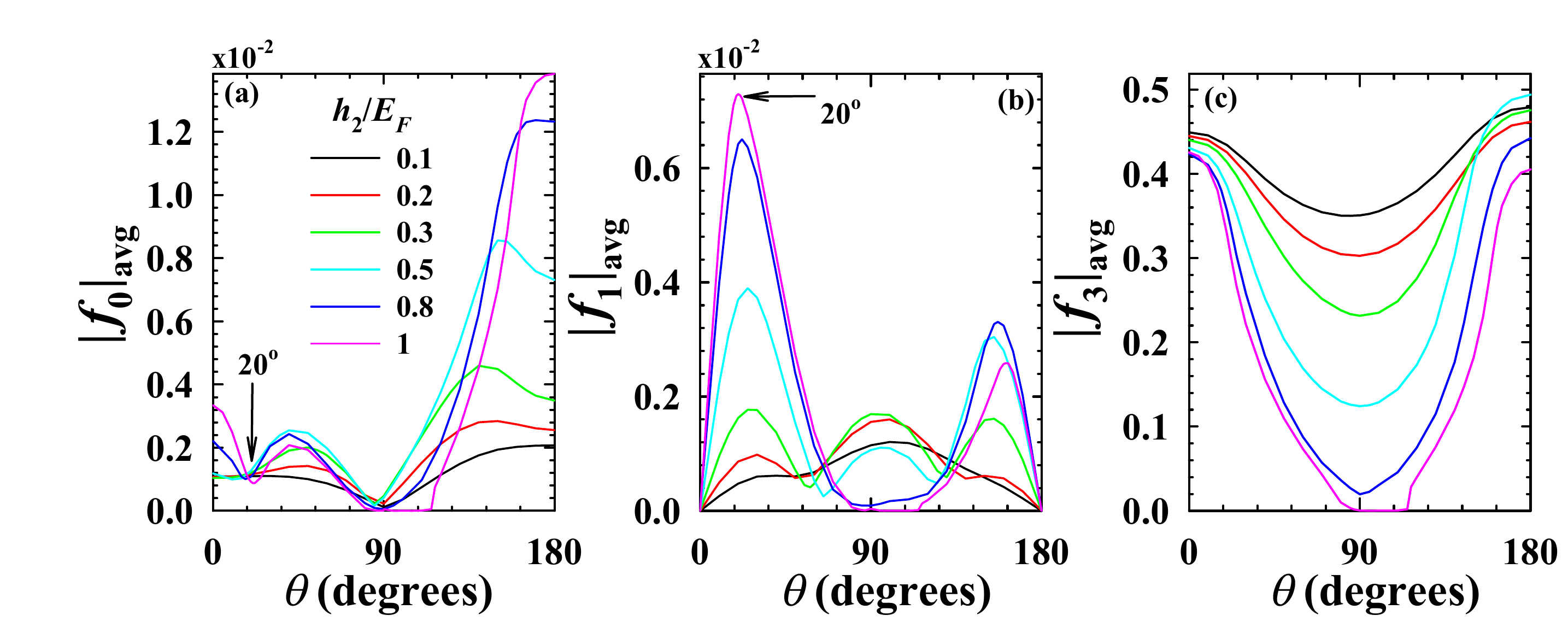}
\includegraphics[width=0.99\columnwidth]{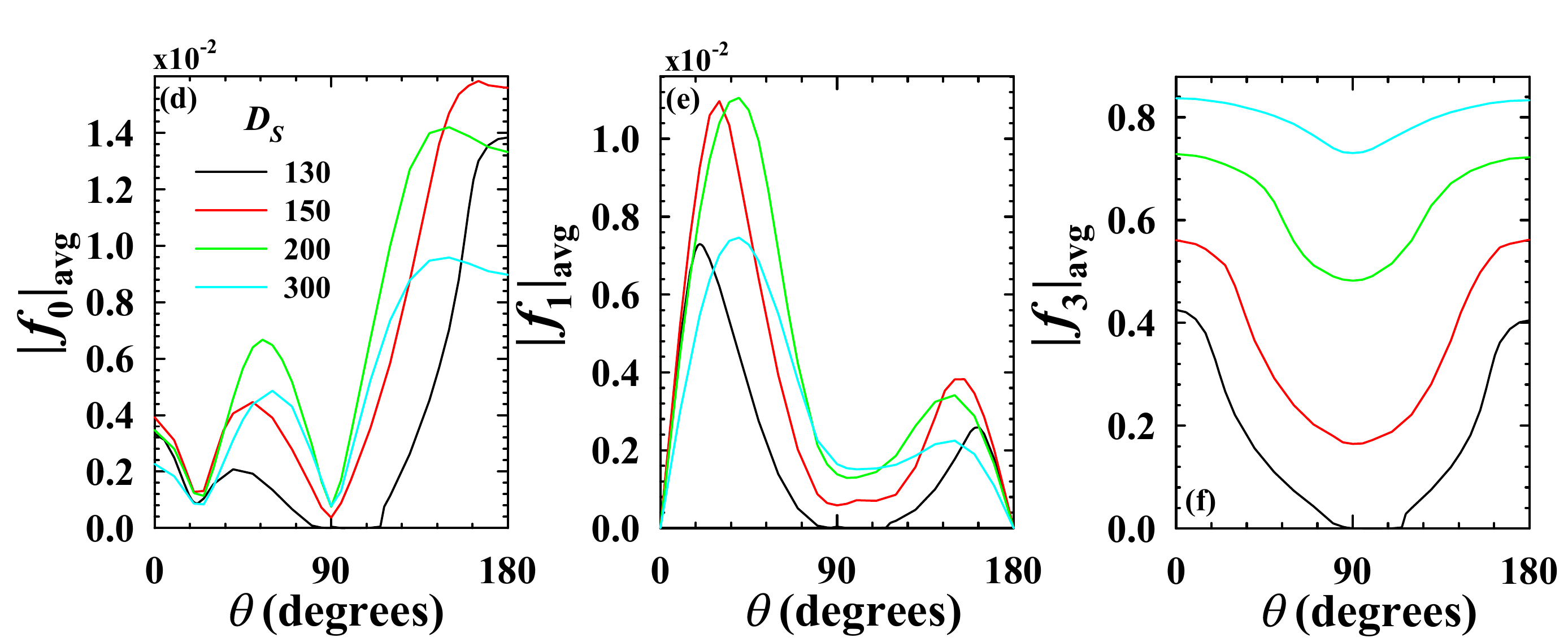}
\caption{
(Color online). The magnitudes of the normalized 
triplet ($f_0, f_1$) and singlet ($f_3$)components 
are shown averaged over the $S$ region and plotted as a function of
 the relative magnetization angle $\theta$.
The temperature is set at $T=0.05 T_0$. 
The top panels (a)-(c) depict 
 differing values of the
exchange field in the $F_2$ region as shown.
All other system parameters are the same as those used in Fig.~\ref{tc_h}.
Panels (d)-(f) correspond to $F_2$ with an 
optimal exchange field of  $h_2/E_F=1$,
and various $S$ widths, as labeled.
}
\label{trip_h}
\end{figure}

\subsection{Critical Temperature and Triplet Correlations}
We first study the critical temperature of the spin valve system.
The linearized
self-consistency expression near $T_c$  takes the form,
$\Delta_i=\sum_q {\cal G}_{iq}\Delta_q$, 
where $\Delta_i$ are the expansion coefficients
for $\Delta(x)$ in the chosen basis.
The ${\cal G}_{iq}$ are the corresponding matrix elements,
which involve sums
of the normal state energies and wavefunctions.
To determine $T_c$, we  compute the eigenvalues 
$\lambda$, of the corresponding eigensystem  
${\bm \Delta}=\lambda {\cal G} {\bm \Delta}$.
When  $\lambda>1$ at a given temperature,
the system is in the superconducting state.
Many of the computational details can be found in Ref.~\onlinecite{ilya}, 
and are omitted here.

It was experimentally observed \cite{singh} 
that a $S F_1 F_2$ spin valve
is most effective at converting singlet Cooper 
pairs to 
spin polarized triplet pairs when $F_2$ is in a
half-metallic phase. To examine  this theoretically, 
we investigate the critical temperature and
corresponding triplet pair generation as a function of $h_2/E_F$ and $\theta$
($h_1/E_F=0.1$ remains fixed).
The width of the
superconducting layer is maintained at $D_S=130$, and the nonmagnetic insert
has a set width corresponding to $D_N=5$. 
The exchange field
$h_2$ varies from $0.1E_F$ to $E_F$ where $h_2=E_F$ corresponds to 
the situation where only one
spin species exists in this region (i.e. the half-metallic phase). 
As seen in
Fig.~\ref{tc_h}, $T_c$ is nearly constant over the full range of $\theta$ when
both ferromagnets are of the same type, i.e.,
when $h_2/E_F=0.1$.
Upon increasing $h_2$ towards the half-metallic limit,
it is apparent that
the spin valve effect becomes
dramatically enhanced, whereby
 rapid  changes in $T_c$ occur when varying $\theta$.
This result therefore clearly supports 
 the assertion that the
use of a half-metal generates
the most optimal spin-valve effectiveness \cite{singh}. 
Large variations in $T_c$ have also been 
found using a diffusive  
quasiclassical approach involving
$S F_1 F_2$ heterostructures lacking the 
normal layer insert \cite{Mironov,fomin}.
%khx3 -----
When 
comparing  $T_c$ in the two collinear
magnetic orientations, 
the 
self-consistently calculated critical temperatures
in Fig.~\ref{tc_h} reveal that the parallel state ($\theta=0^\circ$) has
a smaller $T_c$ compared to
the antiparallel state ($\theta=180^\circ$)
for moderate exchange field strengths. 
For these cases, the two magnets 
can counter one another, leading to a reduction
of their effective pair-breaking effects. This creates
a 
more favorable 
situation  for the superconducting state, causing  $T_c$ to be larger.
The situation reverses
for stronger magnets with $h \gtrsim 0.8$,
and the maximum $T_c$  now arises for  parallel relative orientations
of the magnetizations.  
In between the parallel and antiparallel states, 
$T_c$ undergoes a minimum
that
occurs not at the orthogonal orientation ($\theta=90^\circ$),
but  slightly away from it.
This behavior 
has been observed in ballistic  \cite{wu} and
diffusive  \cite{fomin}  systems where
the minimum in $T_c$ arises from the
leakage of Cooper pairs
that are coupled to the outer $F$ layer
via the generation of the triplet component $f_1$
that is largest near $\theta=90^\circ$.
%khx3 ----

To demonstrate the correlation 
between the strong 
$T_c$ variations 
and the generation of
triplet and singlet 
pairs, Fig.~\ref{trip_h} 
shows 
the magnitudes 
of  the 
equal-spin triplet amplitudes ($f_1$),   opposite-spin triplet amplitudes ($f_0$),
and
the singlet pair amplitudes ($f_3$), each 
averaged over the $S$ region.
For the triplet correlations, 
a representative value for the normalized relative time $\tau$ is set at
$\tau\equiv \omega_D t = 4$.
When the ferromagnet ($F_2$) possesses a large exchange field,
and the relative magnetization angle between $F_1$ and $F_2$ approaches 
an orthogonal state, superconductivity becomes 
severely weakened. Indeed, as Fig.~{\ref{tc_h} demonstrated, 
the singlet pair correlations can become completely destroyed at 
low temperatures ($T\simeq 0.05$), and
orientations in the vicinity of
$\theta\simeq90^\circ$, whereby the system
has transitioned to a normal resistive state.
This is consistent with Fig.~\ref{trip_h}(c), where
the $f_3$ amplitudes 
vanish in the neighborhood of $\theta\approx90^\circ$ and $h_2/E_F = 1$.
As  Fig.~\ref{trip_h}(a) and (b)
illustrates,  the triplet amplitudes also 
 vanish due to the
absence of singlet correlations at those orientations. 
For weaker magnets however, 
the superconducting state 
 never transitions to a normal resistive state over the entire range of $\theta$, 
and 
the well known situation arises whereby
the equal-spin triplet
pairs are  largest for
orthogonal magnetization 
configurations, 
i.e., when the misalignment angle is greatest  ($\theta\simeq90^\circ$).
In all cases however,
the $f_1$ components must always
vanish at $\theta=0$ and $\theta=180^\circ$,
where the relative collinear magnetization alignments are either in the parallel or antiparallel state
respectively. 
It is clear from Figs.~\ref{trip_h}(a) and \ref{trip_h}(b)
that the average behavior of $|f_0|$ and $|f_1|$
 exhibits their most extreme 
 values when  $T_c$  undergoes its steepest variations around
$\theta \approx 20^\circ$ [see Fig.~{\ref{tc_h}].
In particular,
at the half-metallic phase, $f_1$ is greatly enhanced 
while $f_0$ is dramatically
suppressed. 
Therefore,
the considerable variations in $T_c$
is correlated with the
fact that $100\%$ spin-polarized compounds such as ${\rm CrO_2}$ 
result in the optimal generation of spin
triplet correlations \cite{singh}.  
The suppression of $f_0$ at $\theta \approx 20^\circ$
is fairly robust to changes in 
the size of the $S$ region.
As the bottom panels in Figs.~\ref{trip_h}  illustrate,
increasing $D_S$ by several coherence lengths 
causes very little change in the location of the first minimum in
$f_0$ at $\theta\approx 20^\circ$.
The angle $\theta$ that corresponds  to a peak in $f_1$ however,
noticeably shifts  to larger $\theta$, so that at $\theta \approx 20^\circ$, 
$f_1$ is no longer at its peak value.
Therefore, the thinnest $S$ layer width considered
here, $D_S=130$, leads to the most favorable conditions for the
generation of $f_1$ triplet pairs in the superconductor and
 limited coexistence with the $f_0$ triplet correlations.

\begin{figure}
\centering
\includegraphics[width=0.99\columnwidth]{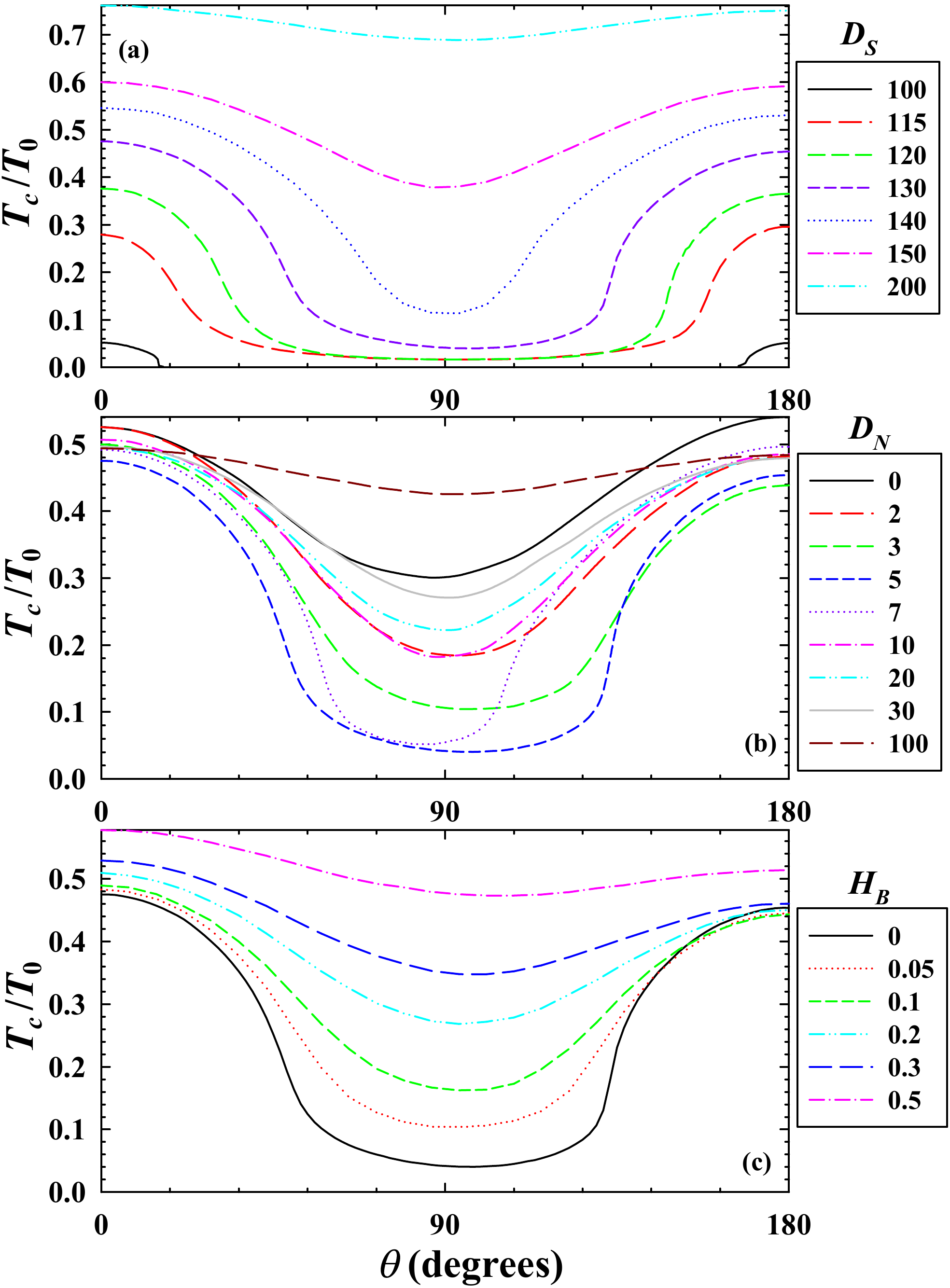}
\caption{
(Color online). Critical temperature $T_c$ as a function of the relative exchange field orientation angle $\theta$.
In (a) the normal metal insert has a width of $D_N=5$, and the $S$ width varies as shown in the legend, 
from $D_S$=100  to $D_S=200$.
In (b) the $S$ width is fixed at $D_S=130$, while the $N$ spacer is varied.
In (c) the effects of interfacial scattering are examined, with $D_S=130$, $D_N=5$.
The legend depicts the various scattering strengths $H_B$ considered.
}
\label{tc}
\end{figure}

Next, Fig.~\ref{tc} shows $T_c$ as a function of the out-of-plane
misalignment angle $\theta$ for differing  (a)  superconductor widths  $D_S$,
(b)  normal layer
widths $D_N$, and (c) spin-independent
interface
scattering strengths $H_B$. 
If the
relative magnetizations were to rotate in-plane,
the $T_c$  behavior discussed here
would be identical,
thus providing additional experimental options for
observing the predicted effects.
In (a), the  sensitivity of $T_c$ to the $S$ layer width is shown. 
The importance of having thin $S$ layers
with  $d_S \sim \xi_0$ (100 in our units) is clearly seen.
In essence,
extremely narrow 
$S$ boundaries restrict  
Cooper pair formation, 
causing the ordered superconducting state to 
effectively become
more ``fragile",
consistent with other 
$F/S$ systems containing thin $S$ layers \cite{wu}.
Indeed, for the thinnest case,
$D_S=100$, 
superconductivity 
completely vanishes 
for most 
magnetization configurations,   
except when  $\theta$ is near
the parallel or antiparallel orientations. 
At the thickest  $D_S$ shown ($D_S=200$),  the  sensitivity to $\theta$  has dramatically
diminished, as pair-breaking effects from the adjacent ferromagnet 
now have a limited overall effect
in the larger superconductor.
For all $S$ widths considered,
the minimum in $T_c$ 
occurs when $\theta$
lies slightly off the orthogonal  
configuration ($\theta=90^\circ$),
consistent with
some quasiclassical 
systems \cite{fomin}.
Next,
in Fig.~\ref{tc}(b)
the $S$ layer thickness is set 
to $D_S=130$, 
while several  nonmagnetic 
$N$ metal spacer widths
are considered.
The presence of the $N$ layer clearly 
plays a crucial
role in the thermodynamics of 
 the spin valve. Indeed,  an optimum $D_N\approx 5$ exists which yields the greatest
 $\Delta T_c (\theta)$:
Increasing or decreasing $D_N$ around this value
can significantly
reduce the size
of the spin valve effect.
Physically, 
this behavior 
is related to the 
spin-triplet conversion 
that takes place
in the ferromagnets
and corresponding enhancement of the equal-spin triplet correlations
in the $N$ layer. This will be discussed in greater detail below. 
For
$D_N$  
much 
larger than the optimal
width,
a severe reduction in 
magnetic interlayer coupling occurs  and 
$T_c$ exhibits little variation with $\theta$.
Finally, 
in Fig.~\ref{tc}(c),
we 
incorporate spin-independent  scattering at each of the spin valve 
interfaces.
A wide range of   
scattering strengths are considered.
We assume $H_j\equiv H \,(j=1,2,3)$, so that
interface
scattering  can be written solely in terms of the
dimensionless parameter
$H_B=  H/v_F$.  
Overall,
the general features and trends for $T_c$
seen previously  
are
retained.
With moderate amounts of interface scattering, 
$H_B=0.1$, we find $\Delta T_c\equiv T_c(\theta=0^\circ)-
T_c(\theta=90^\circ) \approx 0.3 T_0$.
It is immediately evident that 
samples must have interfaces as transparent as possible \cite{singh,klaus_zep}:
the variations in $T_c$ 
with $\theta$
become severely reduced
with increasing $H_B$,
as the phase coherence of the superconducting correlations becomes destroyed. 
%khx4 ---
In all cases, we observe some degree of
asymmetry in $T_c$ as a function of $\theta$,
similar to what has been reported in both diffusive \cite{fomin} and clean \cite{wu} spin valves lacking half-metallic elements.
If it is assumed that the 
band splitting  in $F_2$ is
sufficiently large so that only one spin species can exist, 
%and therefore,only the equal-spin component of the Green function is kept in the calculations, 
a quasiclassical approach has shown
that $T_c$  becomes symmetric with respect to $\theta$ %khp
in the diffusive regime \cite{Mironov}.
% khx4 ---

\begin{figure}
\centering
\includegraphics[width=0.95\columnwidth]{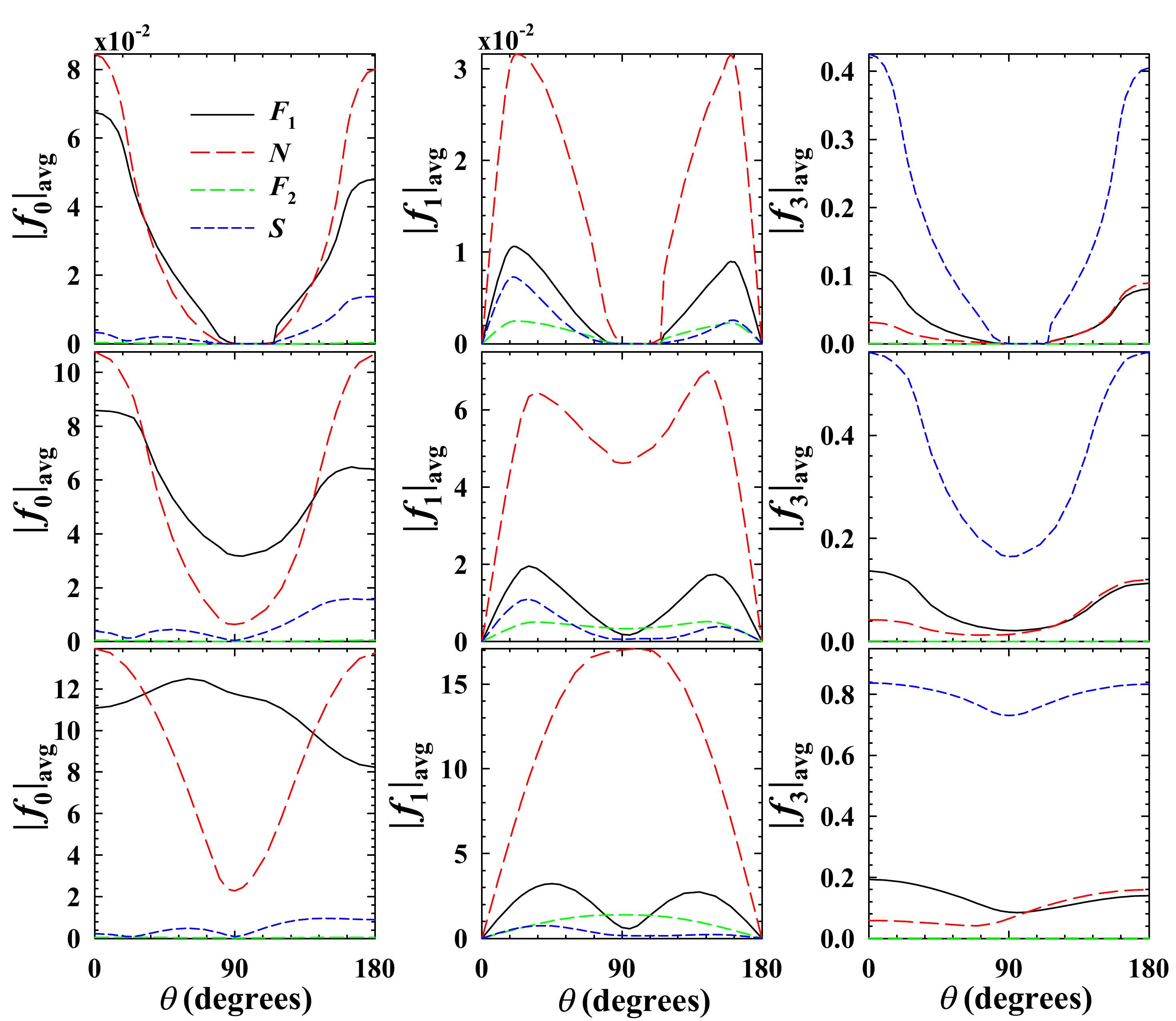}
\caption{
(Color online).
Normalized triplet
($f_0$, $f_1$) and singlet ($f_3$)
amplitudes
versus the relative 
magnetization 
angle $\theta$.
The magnitude
of each pair correlation is
averaged over a given region in the $SF_1NF_2$ 
spin valve, as identified in the legend.
The top, middle, and bottom 
rows correspond to $D_S=130$,
$D_S=150$, and
$D_S=300$ respectively. 
}
\label{triplets}
\end{figure}

To correlate the large spin-valve 
effect observed in Fig.~\ref{tc}  
with the odd-time triplet correlations, we employ
the expressions in Eqs.~({\ref{f0defa}) and (\ref{f1defa}), which describe the spatial
and temporal behavior of the  triplet amplitudes. 
We normalize 
the triplet correlations, 
computed in the low $T$ limit, to
the value of the singlet pair amplitude 
in the bulk $S$. The 
normalized averages of $|f_0|$ and $|f_1|$
are plotted as  functions of $\theta$
in Fig.~\ref{triplets},  at a dimensionless characteristic
time of $\tau = 4$. 
For comparison purposes,  
the singlet pair correlations, 
$f_3$, are also shown
(third column).
In each panel, 
spatial averages over different segments of the spin valve are
 displayed as separate curves (see caption). 
 Each row of figures
 corresponds to different 
 $D_S$: $D_S=130,150,300$ (from top to bottom).
One of the most striking observations is the effect of the normal metal
spacer, which contains a substantial portion of the equal-spin triplet pairs. 
We will see below that the $f_1$ triplet correlations
within the normal metal tend to  propagate into
the adjacent regions of the spin valve as time evolves.
Examining the top two panels of Fig.~\ref{triplets}, the equal-spin  $f_1$ triplet component  in  $S$
clearly dominates 
its opposite spin counterpart when $\theta\approx 20 ^\circ$.
Thus, only  slight deviations from the parallel state ($\theta=0^\circ$)
generates 
triplet correlations within $S$ 
that have spin projection $m=\pm1$.
For each $D_S$ case studied, the singlet $f_3$ amplitudes are 
clearly largest in the $S$ region
where they originate, 
and then decline further in each subsequent  segment.
It is evident also that the $f_1$ triplet pair amplitudes
are anticorrelated to $T_c$ (governed by the behavior of the singlet amplitudes),
which indicates a singlet-triplet conversion process.
\begin{figure}
\centering
\includegraphics[width=0.95\columnwidth]{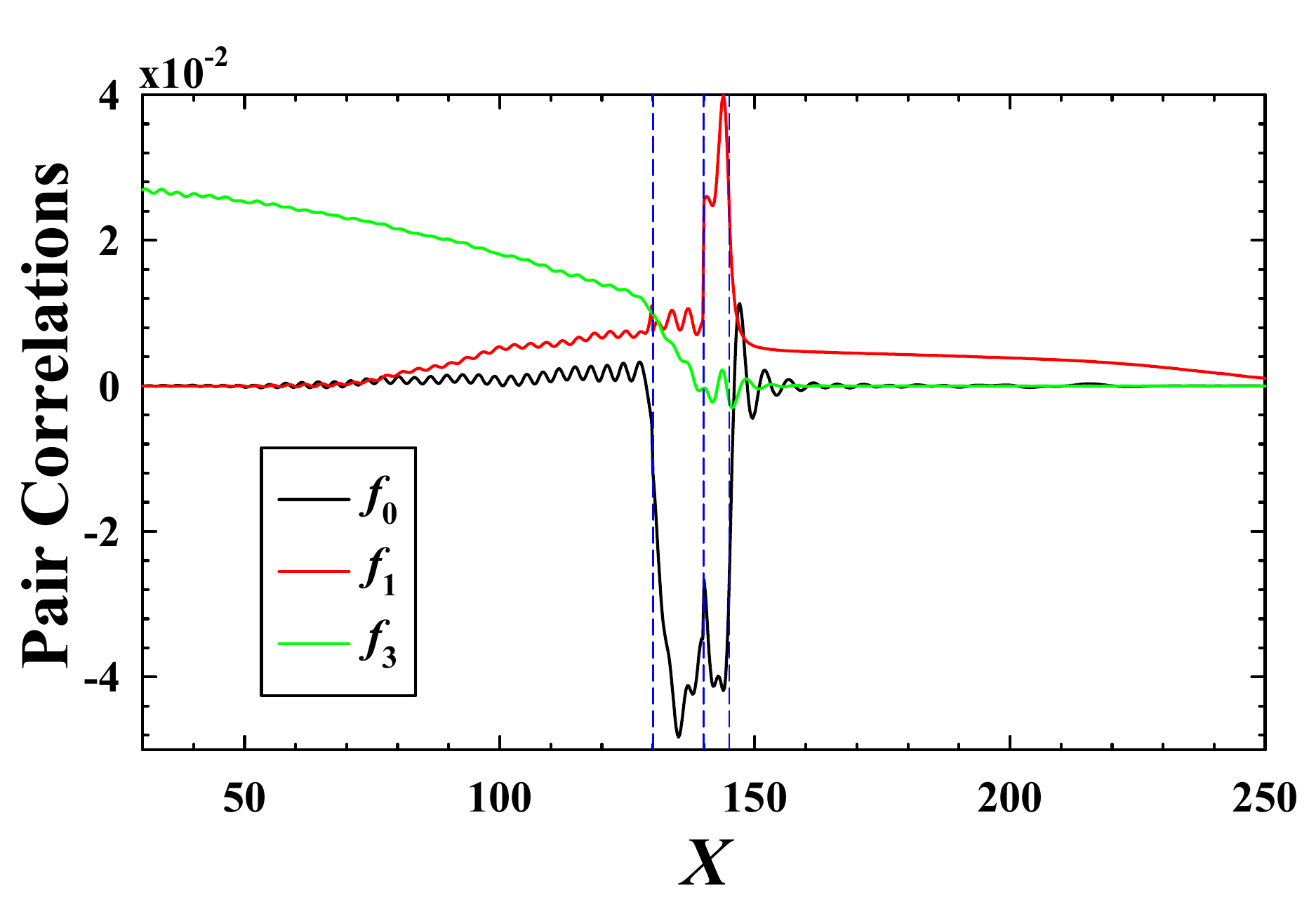}
\caption{
(Color online).
Normalized triplet
($f_0$, $f_1$) and singlet ($f_3$)
amplitudes
versus the dimensionless coordinate $X$.
The relative magnetization orientation is
set to $\theta=20^\circ$. The dashed vertical lines identify the locations of
the interfaces for the $S F_1 N F_2$ structure.
Each segment corresponds to the following ranges:
$X<130$ ($S$ region), 
$130\leq X \leq 140$ ($F_1$ region),
$140 < X \leq 145$ ($N$ region),
and $X > 145$ ($F_2$ region).
The singlet component has been reduced by a factor of 10
for comparison purposes. 
}
\label{pair_spatial}
\end{figure}
Therefore  as more singlet
superconductivity leaks into the ferromagnet side, $T_c$ is
suppressed, and triplet superconductivity is enhanced.  
It is evident that both triplet components vanish around $\theta=90^\circ$,
as was also observed in Fig.~\ref{trip_h}.
This is due to the highly sensitive nature of the gapless superconducting state
that arises in thin $S$ systems, 
whereby the 
singlet 
pair correlations become  rapidly  destroyed as the magnetization vector in $F_1$
approaches the orthogonal configuration. 
Increasing  the size of the superconductor
causes the superconducting state to become more robust to
 changes in $\theta$, and consequently
the system no longer transitions to a resistive state
at $\theta \approx 90^\circ$.
The triplet correlations  reflect this aspect as seen in the middle and bottom panels of 
Fig.~\ref{triplets}, whereby both triplet components have finite values
for the orthogonal orientation.
Overall, there is a dramatic 
change in both triplet components when the $S$ part of the spin valve
is increased in size.
For example, the $f_1$ triplet correlations in $N$
and in $F_2$ evolve from having two peaks
two a single maximum at $\theta = 90^\circ$.
%khx2 ------
The $D_S$ trends also reflect the importance of self-consistency of the pair potential $\Delta(x)$ 
for thinner superconductors, where a self-consistent singlet component  $f_3(x)$
can substantially decline, or vanish altogether, in contrast to simple step function.
Indeed, the  observed
 disappearance   of the singlet and triplet
 pair correlations
 for thin superconductors at $\theta\simeq 90^\circ$  (see top panels),
can only occur if
the pair potential is calculated self-consistently [Eq.~(\ref{del})],
thus ensuring that the free energy of the system is lowest \cite{gennes}.
As will be seen below,
this important step permits the proper description of the proximity effects leading to 
nontrivial spatial behavior of $\Delta(x)$ in and around
the interfaces for both the superconductor and ferromagnets \cite{klaus_first}.
In common non self-consistent approaches, where $\Delta(x)$ is treated
phenomenologically as a  prescribed constant 
in the $S$ region, this vital behavior is lost.
%khx2 ------

Next, in Fig.~\ref{pair_spatial} we present
the spatial behavior of the 
real parts of the
triplet and singlet 
pair correlations
throughout each segment of the spin valve.
We choose $\theta=20^\circ$ in order to optimize the
$f_1$ triplet component in $S$.
The other parameters used correspond to $D_S=130$,
$D_N=5$, and $T=0.05$.
Proximity effects are seen to result in a reduction
of the singlet $f_3$ correlations in the $S$ region
near the interface at $X=130$. As usual, this decay occurs over the
coherence length $\xi_0$. The singlet amplitude then declines within
the $F_1$ region before  undergoing oscillations and 
quickly dampening out in the half-metal. Thus, as expected, 
the singlet Cooper pairs cannot be sustained in the half-metallic segment
where only one spin species exists.
Within the half-metal, the triplet component,
$f_0$ (also comprised of opposite-spin pairs),
undergoes damped oscillations  similar to
the $f_3$ correlations. It is notable that the 
triplet $f_0$ component 
is severely limited in the $S$ region, in stark contrast to the singlet correlations. Therefore,
 the $f_0$ correlations in this situation
 are  confined mainly to the $F_1$ and $N$ regions.
The equal-spin $f_1$ triplet component on the other hand, is seen to
pervade every segment of the spin valve: The $f_1$ correlations are 
enhanced in the $N$ region, similar 
in magnitude to $f_0$, but then exhibit a slow decay in both the $S$ and half-metallic regions.

\begin{figure}
\centering
\includegraphics[width=0.9\columnwidth]{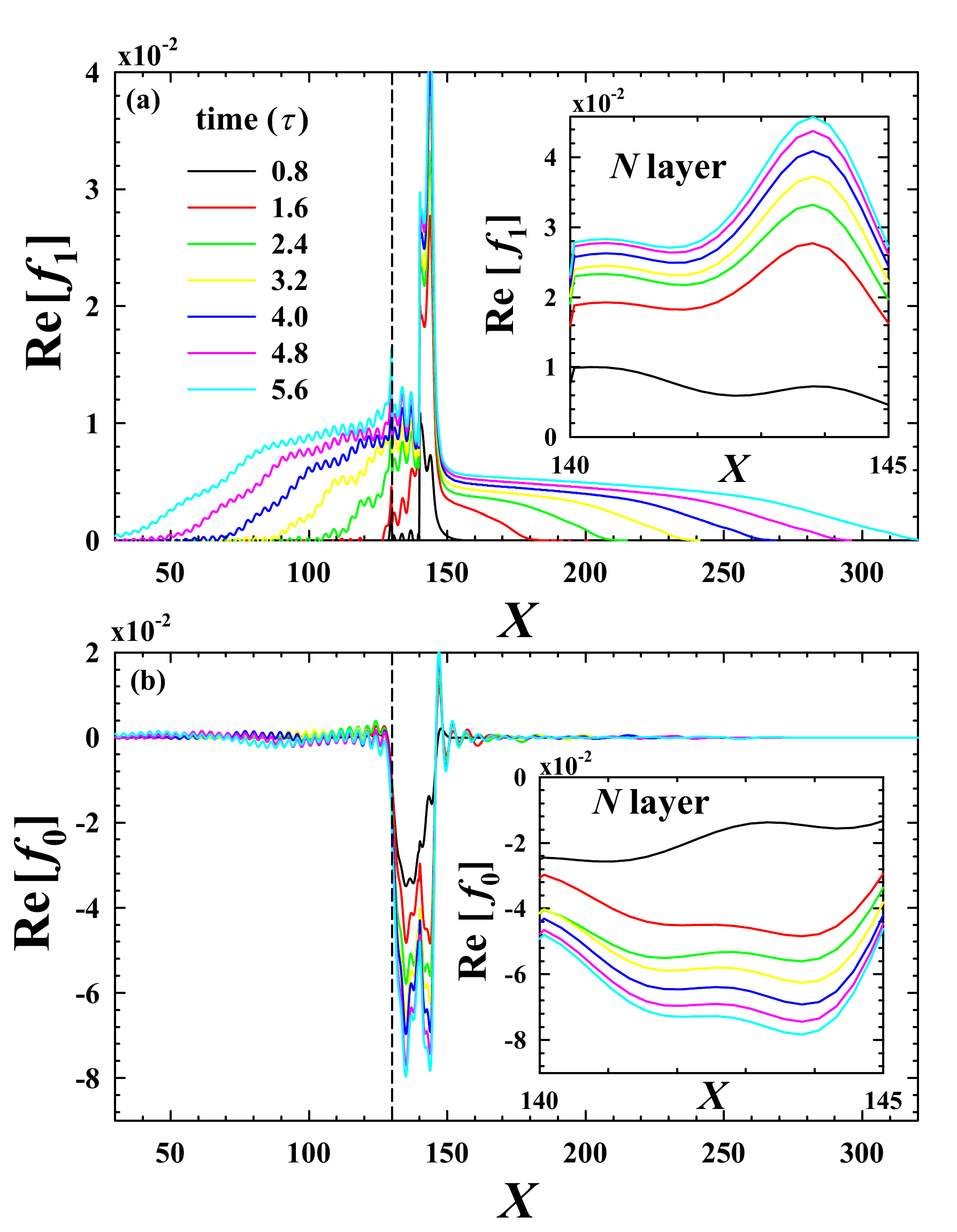}
\caption{
(Color online).
Time evolution of the localized spatial dependence of the $f_0$ and $f_1$
triplet correlations. The insets depict magnifications of the $N$ regions 
($140\leq X \leq145$).
The dimensionless time parameter $\tau\equiv \omega_D t$ varies from
$0.8$ to $5.6$ in increments of $0.8$. Initially, the $f_1$ component
predominately populates the $N$ region,
and then progressively moves outward into each segment
of the spin valve with increasing time. 
The $f_0$ component initially occupies the $F_1$ and $N$ layers,
and then remains confined to those regions at higher $\tau$.
Each dashed vertical line identifies the $S$ interface.
}
\label{triplets_time}
\end{figure}
To further clarify the role of the triplet correlations  in the spin valve,
we now discuss the explicit relative time evolution
of the  triplet states
in Fig.~\ref{triplets_time}. Snapshots of the real parts of the 
triplet amplitudes are shown in equal increments of the relative time parameter $\tau$.
 The angle $\theta$ is fixed at $\theta=20^\circ$, again corresponding to when
the triplet correlations with $m=\pm1$
projection of the $z$-component of the total spin  in the superconductor is largest (see Fig.~\ref{triplets}).
 The spatial range shown permits
visualization of both triplet components throughout much of the system.
Starting at the earliest time $\tau=0.8$, we find that $f_1$ 
mainly  populates the nonmagnetic $N$ region,
and then as $\tau$ increases, propagates into the $F_1$ and $F_2$ regions before extending into 
the superconductor (left of the dashed vertical line). 
Meanwhile, $f_0$ is essentially confined to the $F_1$ and $N$ regions,
with limited presence in
the $S$ and $F_2$ layers.
Since the characteristic  length $\xi_F$ over which the $f_0$ correlations
 modulate in $F_2$ is inversely proportional to $h_2$,
$f_0$ declines
sharply in the half-metallic  region. 
Also, in agreement with Fig.~\ref{triplets}, for $\theta=20^\circ$ and $D_S=130$, there is also
 a limited presence of $f_0$ in the superconductor.
 The superconductor therefore has  $|f_1| \gg |f_0|$, 
which by using the appropriate experimental probe, can  reveal 
signatures detailing the presence of equal-spin pairs $f_1$ \cite{bernard}.

\begin{figure}
\centering
\includegraphics[width=\columnwidth]{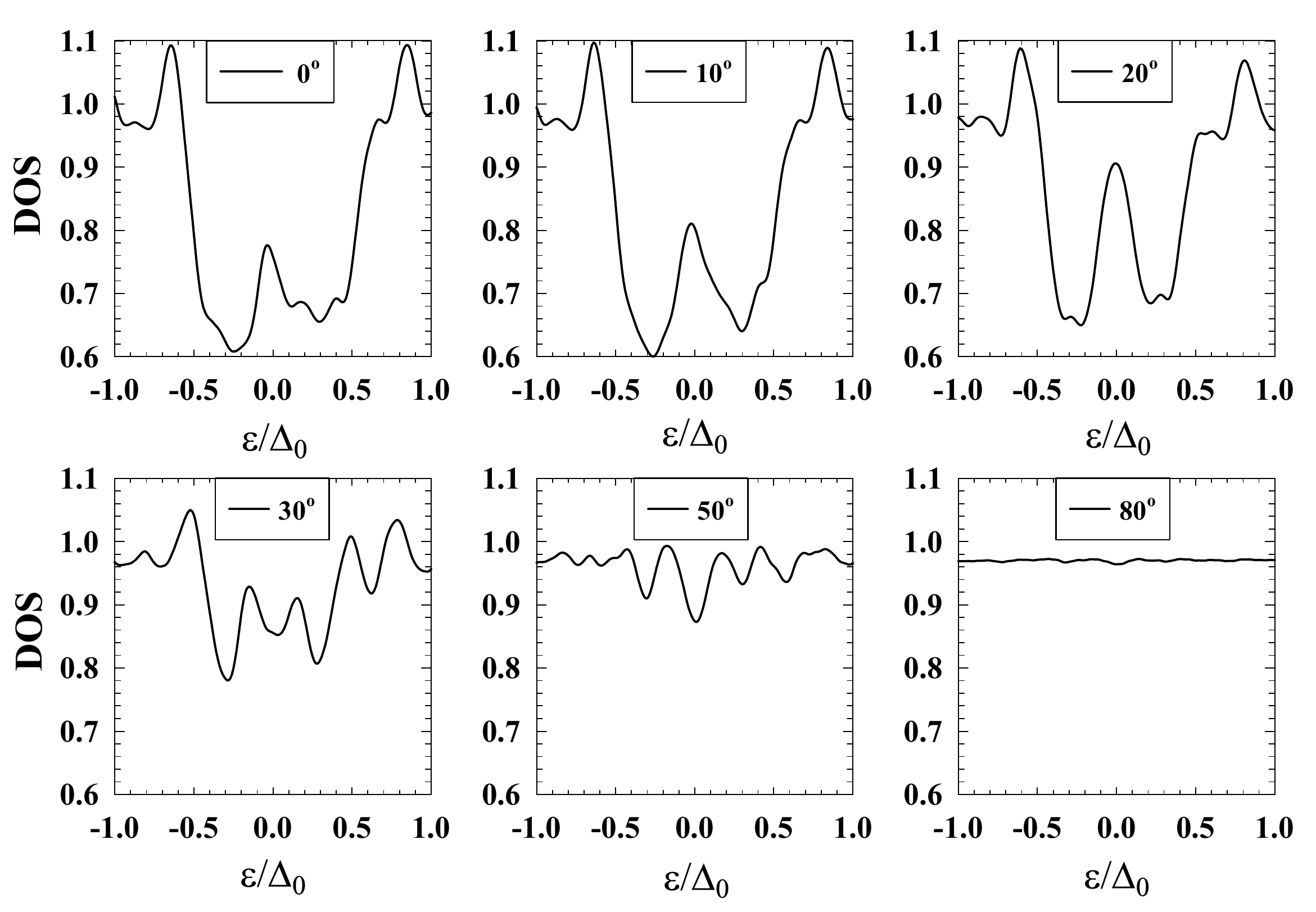}
\caption{
(Color online). Signatures of equal-spin triplet correlations:
The normalized local DOS in the superconductor  for various relative  
magnetization orientations,
$\theta$. In the range $0^\circ\leq \theta \leq20^\circ$, the DOS
possesses peaks at zero energy 
which grow until they become inverted at $\theta=30^\circ$. 
The well defined, 
prominent ZEP at $\theta=20^\circ$
corresponds to the maximal generation of equal-spin triplet amplitudes in the $S$ region, 
as shown in Fig.~\ref{triplets}.
}
\label{dos}
\end{figure}

\begin{figure*}
\centering
\includegraphics[width=\textwidth]{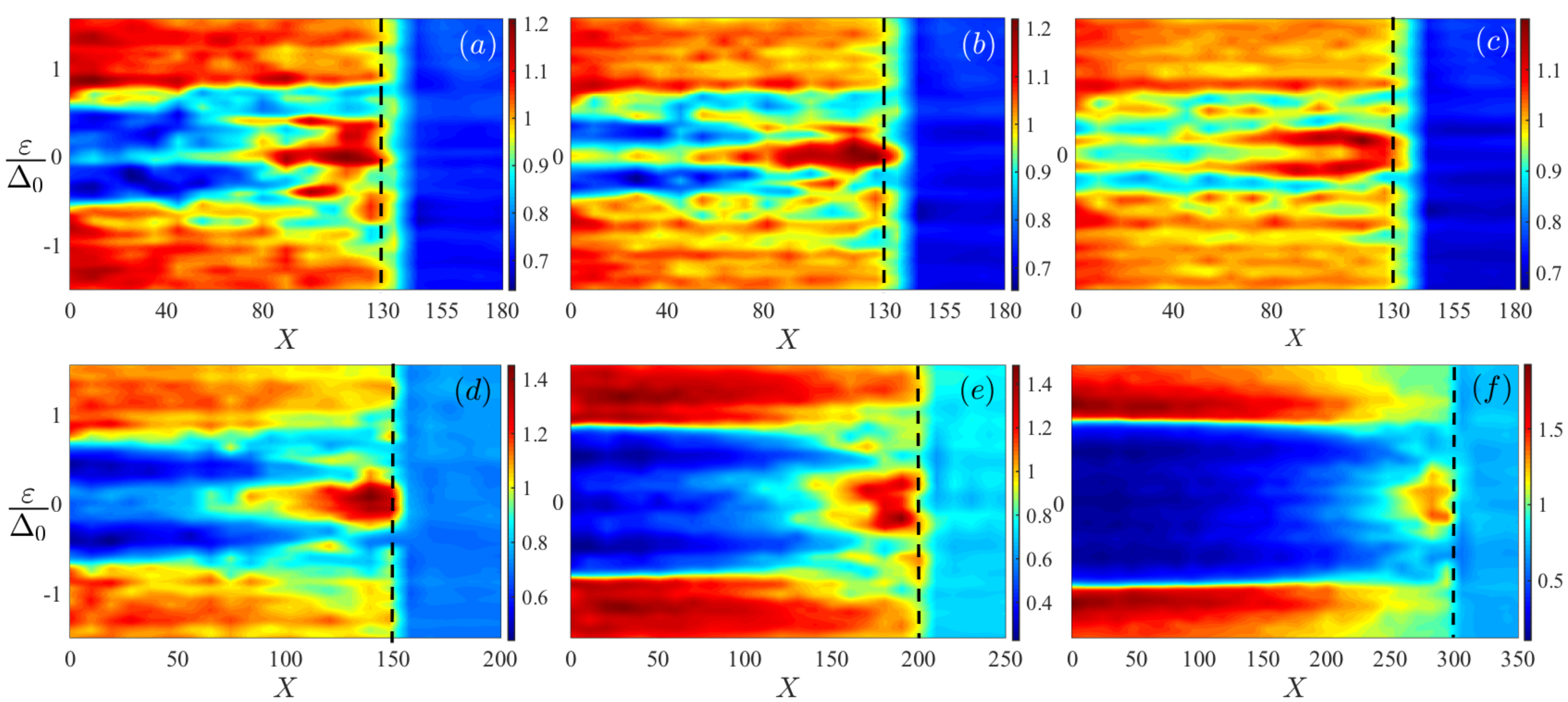}
\caption{
(Color online).   Top panels: 
The normalized
spatially and energy resolved DOS at three
    different orientations of the relative magnetization angle:
  (a)  $\theta=10^\circ$, (b) $\theta=20^\circ$, and (c) $\theta=30^\circ$.
    Panels (a)-(c)  pertain to a single system with a narrow $S$ layer 
    of width  $D_S=130$.
     The spatial region extending from $X=0$ to $130$ therefore 
     corresponds to  the
    superconducting region, and $X>130$ pertains to the 
    remaining 
    layers of the spin valve.  
    Bottom panels:
     the DOS is shown for three
    different $S$ layer thicknesses: (d) $D_S=150$, (e) $D_S=200$,
    and  (f) $D_S=300$, where $\theta$ is now fixed at $20^\circ$.
    The dashed vertical lines identify the interface between $S$ and $F_1$.
}
\label{2ddos}
\end{figure*}

\subsection{Density of States}
To explore these proximity induced
signatures further, 
we investigate
the experimentally relevant local DOS.
An important spectroscopic tool for
 exploring 
proximity
effects on an atomic scale with sub-meV energy resolution
is the scanning tunneling microscope (STM). 
We are interested in determining the local DOS in the outer
$S$ segment  of the  $SF_1 N F_2$ spin valve. 
By positioning a nonmagnetic STM tip at the edge of
the $S$ region, the tunneling current ($I$)  and voltage ($V$)
 characteristics can be measured \cite{bernard}.
 This technique yields a direct probe
of the available electronic states with energy $eV$ near the tip.
The corresponding  differential conductance $dI(V )/dV$ over the
energy range of interest is then proportional to the local DOS.
The vast majority of past works
only considered the DOS in the ferromagnet side where
the $f_1$ correlations were expected
to dominate \cite{bernard,klaus_zep,golu}.
However unavoidable experimental
issues related to noise and thermal broadening can 
yield inconclusive data.
As we have shown above, with the proper alignment of relative magnetizations,
one can  generate  a finite $f_1$ in $S$  accompanied by relatively limited  $f_0$,
thus presenting an opportunity to detect the important
 triplet pairs with spin $s=\pm 1$.
By avoiding comparable admixtures of the two triplet components,
experimental signatures of the equal-spin
 triplet correlations  should be  discernible.
To investigate this further,
the six
panels in Fig.~\ref{dos} 
show the normalized DOS evaluated near
 the edge of the superconductor for a wide variety of 
 orientation angles $\theta$.
All plots are normalized to the corresponding
value in a bulk sample of $S$ material in its normal state. 
As shown, each panel ranges from 
 a mutually parallel  ($\theta=0^\circ$) to
 a  nearly orthogonal magnetization state  ($\theta=80^\circ$).
In each
case considered, we again have  $D_N=5$ and $D_S=130$.
Examining  the top row of panels, 
traces are seen of the well-known BCS peaks that  
have now been shifted to subgap energies
due to  proximity and size effects. 
There also exists bound states at low energies that
arise from quasiparticle interference effects.
By sweeping the angle $\theta$ from the relative parallel
case  ($\theta=0^\circ$) to slightly out of plane  ($\theta=20^\circ$),
the zero energy quasiparticle states 
 become significantly more pronounced. 
This follows from the fact that
 strong magnets
tend to shift the relative magnetizations leading to maximal $f_1$ generation
away
from the expected  
orthogonal alignment at $\theta=90^\circ$ \cite{klaus_zep}.
The top panels  reflect  the
gapless superconducting state  often found in $F/S$ heterostructures \cite{gap},
superimposed with
the triplet induced  zero-energy peaks.
The modifications to the superconducting state
in the form of
a subgap DOS in the superconductor 
is another signature that is 
indicative of the presence of 
spin-triplet pair correlations \cite{bernard}.
Finally, as $\theta$ rotates further
out of plane  ($\theta>20^\circ$),
the former ZEP's become inverted and vanish
when
$\theta=80^\circ$,  exhibiting   a relatively flat DOS 
where  the system has essentially transitioned to the normal state (see Fig.~\ref{tc}).

A complimentary global  view of the above phenomena is presented in
 Fig.~\ref{2ddos}, where both  the spatially and energy resolved DOS
is shown at various $\theta$ (top panels) and $D_S$ (bottom panels). The top panels (a)-(c)
depict the DOS
for the same parameters and normalizations used in Fig.~\ref{dos}, 
and at three  orientations: $\theta=0^\circ, 10^\circ, 20^\circ$.
It is evident that increasing the misalignment
angle $\theta$, causes the ZEP in the $S$ region
to become enhanced, reaching its maximum at
$\theta\approx 20^\circ$. 
At this angle the ZEP extends through 
much of the system, including to a small extent, the $F_2$ side. However,  within
 $S$, the ZEP is clearly more dominate \cite{bernard}. 
For the bottom panels, (d)-(f),
the relative magnetization orientation is fixed at $\theta=20^\circ$,
and three larger $S$ layers are shown:
 $D_S=150$, $D_S=200$, and $D_S=300$.
Increasing the $S$ layer widths illustrates
the ZEP evolution towards a familiar gapped DOS of a BCS form.
As seen, the ZEP is maximal in the
superconducting region near the $S/F_1$ interface. By
increasing $D_S$, the ZEP in the $S$ side becomes diminished  until for 
sufficiently
 large $D_S$, that is, $D_S\approx 200$, the 
 well-known singlet superconducting
gap begins to emerge throughout much of the superconductor. 
At an even larger $D_S$ ($D_S=300$), 
the ZEP has clearly weakened even further.
Finally, for the experiment reported in Ref. \onlinecite{singh}, a
peak in the resistive transitions at external fields of $B>0.25\, {\rm T}$  was
observed immediately before the critical temperature whereby the system
has transitioned to the superconducting phase. This peak in the transition curves was
believed to be caused by the influence of the external field, effectively  creating
 a $SF_1F'F_2$ type of configuration. 
We investigated 
such a configuration for various strengths and orientations of the
$F'$ ferromagnet, and no evidence was found that was
suggestive of anomalous behavior 
near $T_c$ for
 $F'$ with weak exchange fields. 
Note that the system under consideration is 
translationally invariant in the $yz$ plane (see Fig. \ref{schematic}). Therefore, 
the spin valve structure may experience a Fulde Ferrell-Larkin-Ovchinnikov phase during
 its phase transition from the superconducting to normal phase, although in a narrow region of parameter space \cite{loff1,loff2}.

\subsection{Spin Currents}
To reveal further details of the exchange interaction which 
controls the
behavior and type of triplet correlations present in the system,
we next examine the characteristics  of the spin currents that  exist within the spin valve.
When the magnetizations in $F_1$ and $F_2$ are noncollinear,
the exchange interaction in
the ferromagnets creates a
spin current ${\bm S}$ that flows in parts of the system,
even in the absence of a charge current.
If the spin current  varies spatially,
the corresponding nonconserved spin currents 
in $F_1$ and $F_2$ 
generate 
a mutual torque that tends to rotate the magnetizations of
the two ferromagnets. 
This process is embodied in the spin-torque continuity equation \cite{joe1,joe2}
which describes the time evolution of the spin density $\bm \eta$:
\begin{align}
\frac{\partial}{\partial t} \langle \eta_i (x)\rangle + \frac{\partial}{\partial x} S_i(x) = \tau_i(x), \quad i=x,y,z,
\label{cont}
\end{align}
where $\bm \tau(x)$ is the 
 spin transfer
torque (STT): ${\bm \tau}(x) = -(2/\mu_B) {\bm m} (x) \times {\bm h}(x)$, ${\bm m}(x)$ is the magnetization,
and $\mu_B$ is the Bohr magneton (see Appendix~\ref{appA}).
The spin current tensor here has been reduced to vector form
due to the quasi-one dimensional nature of the geometry.
We calculate 
  ${\bm S}(x)$ 
  by performing the appropriate sums
  of quasiparticle amplitudes and energies  [see Eq.~(\ref{spinall})].
 In
the steady state, the continuity equation, Eq.~(\ref{cont}), determines 
 the torque  by simply evaluating
the derivative of the spin current as a function of position: 
 ${ \tau}_i(x) = \partial {S}_i(x)/\partial x$.
The net torque acting within the boundaries of e.g., the $F_1$ layer,
 is therefore 
 the change in spin current across the two interfaces bounding that region:
\begin{align} \label{tnet}
S_y(d_S+d_{F_1})-S_y(d_S)=\int_{F_1} dx \tau_y.
\end{align}
In equilibrium,  
the net $\tau_y$ in $F_2$ is opposite to its counterpart in $F_1$.
Since no spin current flows in the superconductor, we have
$S_y(d_S)=0$,  and the net torque in $F_1$
is equivalent to the spin current flowing through $N$. 

In our setup, the exchange field in $F_1$ is
directed 
 in the $x-z$ plane,
and therefore  the spin current
and torque are directed orthogonal to this plane (along the interfaces in the $y$ direction).
Likewise, if the magnetizations were varied in the $y-z$ plane, the spin currents would be directed along $x$.
Figure~\ref{spin} thus  illustrates the normalized spin current $S_y$ as a function of
the dimensionless position $X$. The normalization factor  $S_0$
is written in terms of  $n_e v_F$, where  $n_e=k_F^3/(3\pi^2)$,
and $v_F=k_F/m$.
Several equally spaced magnetization orientations $\theta$ are considered, ranging from 
parallel  ($\theta=0^\circ$), to orthogonal ($\theta=90^\circ$).
Within the two $F$ regions, $S_y$ tends to undergo damped oscillations,
while in $N$ there is no exchange interaction (${\bm h}=0$), and consequently the spin current is constant for a given $\theta$.
The main plot shows that when $\theta=0^\circ$, 
$S_y$ vanishes
throughout the entire system, as expected
for  parallel magnetizations. 
By varying $\theta$, spin currents are induced due to the misaligned magnetic moments in the $F$ layers.
If the exchange field is rotated
slightly out of plane, such that  $\theta \lesssim 30^\circ$, 
it
generates on average, negative
spin currents in the $N$ and $F_1$ regions.
As shown, these spin currents 
 reverse their polarization direction for larger $\theta$. 
 This
behavior is consistent with the inset, 
which shows  how tuning $\theta$  affects $S_y$ (or equivalently, the net torque) in $N$.
Thus, by manipulating $\theta$,  the strength and direction of  the spin current in the normal metal
can be controlled, or even eliminated completely  
 at  $\theta\approx 34^\circ$.
By varying $\theta$ about this angle, the
overall torque, which tends to align the magnets in a particular direction,
can then reverse in a given magnet.
For $\theta \approx 15^\circ$ and $\theta \approx 160^\circ$,  the 
inset also clearly shows an enhancement of the magnitude of the spin currents,
which coincides  approximately 
to the orientations leading to an  increase
in the spin-polarized triplet pairs observed in Fig.~\ref{triplets}.

\begin{figure}[t!]
\centering
\includegraphics[width=\columnwidth]{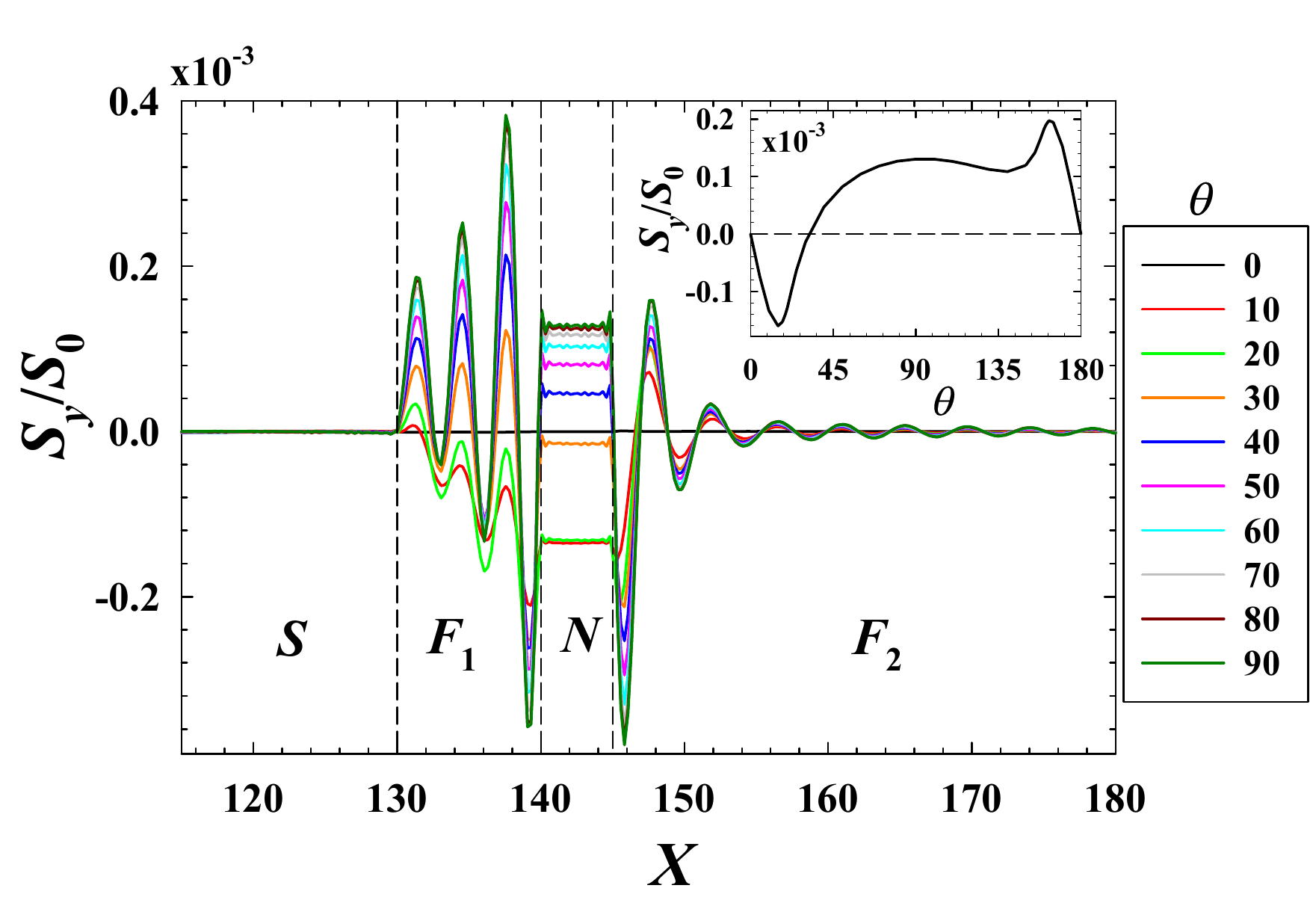}
\caption{
(Color online). Spin current $S_y$ as a function of position $X$ in the spin valve.  Several
magnetization orientations $\theta$ are considered as shown in the legend. The dashed vertical lines identify the interfaces
of each layer as labeled. The inset corresponds to the spin current within the $N$ region.
}
\label{spin}
\end{figure}
In conclusion, 
motivated by recent experiments \cite{singh,bernard}, a hybrid $S F_1 N F_2 $ spin valve containing  a
half-metallic ferromagnet has been theoretically investigated, revealing  a
sizable spin-valve effect
 for thin superconductors with widths close to $\xi_0$. 
Through self-consistent numerical calculations, 
the contributions from both the 
equal-spin ($f_1$) and opposite-spin ($f_0$) triplet
correlations have been identified  as the relative magnetization angle $\theta$ varies.
We found that when the magnetization in $F_1$ is directed slightly out-of-plane,
the magnitude of $f_1$ in $S$ is maximized, while for $f_0$ it  is very small.
By investigating  the DOS in the superconductor 
over a broad range of  $\theta$,
we  were able to identify   the emergence of
zero energy peaks (ZEPs) in the DOS
that coincide with peaks in  the averaged $|f_1|$. 
Our results show,
to a large extent,  good agreement with experimental
observations as well as the physical origins of these effects.
We have thus established a clear, experimentally identifiable role that  the triplet correlations
play in this new class of half-metallic spin valve structures.
 For future work, it would be interesting to
study  the transport properties of  these types of spin valves 
by investigating  the self-consistent
charge and spin currents as they pertain to dissipationless spintronics applications.

\section{Acknowledgements} 
This work was supported 
 in part by ONR and a grant of HPC resources
from the DOD HPCMP.  We thank N. Birge for
a careful reading of the manuscript and helpful comments. 

\appendix\section{Spin Currents}\label{appA}

In order to calculate  the spin currents flowing within the spin valve,
it is convenient to employ
the Heisenberg picture 
to determine the time evolution of the spin density, 
${\bm \eta}({\bm r},t)$,
\begin{align} \label{scom}
\frac{\partial}{\partial t} \langle {\bm \eta}({\bm r},t) \rangle = i \langle 
[{\cal H}_{\rm eff},{\bm \eta}({\bm r},t)] \rangle,
\end{align}
where ${\bm \eta}({\bm r}) $ is the spin density operator defined as,
\begin{align} \label{spinop}
{\bm \eta}({\bm r})  = \psi^\dagger({\bm r})  {\bsigma} \psi({\bm r}).
\end{align}
We define the
effective BCS Hamiltonian \cite{gennes}, ${\cal H}_{\rm eff}$, via
\begin{align} \label{heff}
{\cal H}_{\rm eff} =&\int d^3 r\Bigl\lbrace \psi^\dagger ({\bm r})[  {\cal
H}_0({\bm r})-  {\bm h}({\bm r}) \cdot {\bm \sigma}] \psi ({\bm r})  \nonumber \\
 &+
 \Delta({\bm r}) \psi^\dagger_\uparrow({\bm r}) \psi^\dagger_\downarrow({\bm r})
+\Delta^*({\bm r}) \psi_\downarrow({\bm r}) \psi_\uparrow({\bm r})
\Bigr\rbrace,
\end{align}
where
$\psi_\sigma^\dagger({\bm r}), \psi_\sigma({\bm r})$  denotes the fermionic  field operators
with spin projections $\sigma=\uparrow,\downarrow$ along a given quantization axis, and $ {\bsigma} $ is the usual vector of Pauli matrices.
Inserting the Hamiltonian, Eq.~(\ref{heff}), into (\ref{scom}) yields the following
 continuity equation:
\begin{align} \label{scon}
\frac{\partial}{\partial t} \langle {\bm \eta}({\bm r},t) \rangle + \frac{\partial {\bm S}}{\partial x} &= 
{\bm \tau},
\end{align}
where ${\bm S}$ is the spin current which  in our 
geometry is a vector (in general it is a tensor).
The spin-transfer torque, ${\bm \tau}$, is given by: 
\begin{align} \label{stt}
{\bm \tau}=-i \langle \psi^\dagger({\bm r}) [{\bm h}\cdot{\bm \sigma},{\bm \sigma}] \psi ({\bm r})\rangle 
=2 \langle \psi^\dagger({\bm r}) [{\bm \sigma}\times{\bm h}] \psi ({\bm r})\rangle.
\end{align}
Recalling the expression 
for the local magnetization, ${\bm m}({\bm r})$,
\begin{align} \label{mag}
{\bm m}({\bm r})  =-\mu_B\, \langle {\bm \eta}({\bm r})   \rangle,
\end{align}
this permits the torque in Eq.~(\ref{stt}) to  be written as,
\begin{equation} \label{tau1}
{\bm \tau}
=2\langle \psi^\dagger({\bm r}) {\bm \sigma}\psi ({\bm r})\rangle\times{\bm h}
= -\frac{2}{\mu_B} {\bm m} \times {\bm h}.
\end{equation}

In the steady state, and when a torque is present,
the spin current therefore must have at least one spatially varying component.
After taking the commutator in Eq.~(\ref{scom}),
the explicit expression for the 
spin-current is found to be,
\begin{align}  
{\bm S} = -\frac{i}{2m} \Bigl \langle \psi^\dagger({\bm r}) {\bm \sigma} \frac{\partial \psi({\bm r})}{\partial x}  
- \frac{\partial  \psi^\dagger({\bm r})}{\partial x} {\bm \sigma} \psi({\bm r})
\Bigr \rangle,
\end{align}
where
 for
our quasi-one-dimensional systems,
  the vector 
${\bm S}$ 
represents the spin current flowing along the $x$ direction
with spin components $(S_x,S_y,S_z)$.
To write the spin current in terms of the calculated quasiparticle amplitudes and energies,
the field operators
are directly expanded by
means of a Bogoliubov transformation \cite{gennes}:
\begin{subequations}
\label{bv}
\begin{align}
\psi_{\uparrow}({\bm r})&=\sum_n \left(u_{n\uparrow}({\bm r})\gamma_n - v^*_{n\uparrow}({\bm r})\gamma_n^\dagger\right), \\
\psi_{\downarrow}({\bm r})&=\sum_n \left(u_{n\downarrow}({\bm r})\gamma_n + v^*_{n\downarrow}({\bm r})\gamma_n^\dagger\right),
\end{align}
\end{subequations}
where $u_{n\sigma}$ and $v_{n\sigma}$ are the quasiparticle and
quasihole amplitudes,
 and
$\gamma_n$ and $\gamma_n^\dagger$ are the Bogoliubov quasiparticle
annihilation and creation operators, respectively. By
directly considering the commutation relations for the quantum mechanical
operators, the following expectation values must be satisfied
throughout our calculations:
$\langle \gamma_n^\dagger \gamma_m \rangle = \delta_{nm} f_n$,
$\langle \gamma_m \gamma^\dagger_n \rangle = \delta_{nm} (1-f_n)$, and
$\langle \gamma_n \gamma_m \rangle =  0$.
Here $f_n$ is the Fermi function
which depends on the temperature $T$
and quasiparticle energy  $\varepsilon_n$: $f_n=(\exp[\varepsilon_n/(2 T)]+1)^{-1}$.
We can now expand each spin component of the spin current in terms of 
the quasiparticle amplitudes to obtain \cite{joe1,joe2}: 
\begin{widetext}
\begin{align} \label{spinall}
&{S}_x =\hspace{-.1cm} 
-\frac{i}{2m}\sum_n \Biggl\lbrace f_n\Bigl[u_{n\uparrow}^* \frac{\partial u_{n \downarrow}}{\partial x}+
u_{n\downarrow}^* \frac{\partial u_{n \uparrow}}{\partial x}-
u_{n\downarrow} \frac{\partial u^*_{n \uparrow}}{\partial x}-
u_{n\uparrow}\frac{\partial u^*_{n \downarrow}}{\partial x} \Bigr ] 
-(1-f_n)
\Bigl[v_{n\uparrow} \frac{\partial v^*_{n \downarrow}}{\partial x}+
v_{n\downarrow} \frac{\partial v^*_{n \uparrow}}{\partial x}-
v^*_{n\uparrow} \frac{\partial v_{n \downarrow}}{\partial x}-
v^*_{n\downarrow} \frac{\partial v_{n \uparrow}}{\partial x} \Bigr ] \Biggr\rbrace,  \\
&{S}_y = \hspace{-.1cm}
-\frac{1}{2m}\sum_n \Biggl\lbrace f_n\Bigl[u_{n\uparrow}^* \frac{\partial u_{n \downarrow}}{\partial x}-
u_{n\downarrow}^* \frac{\partial u_{n \uparrow}}{\partial x}-
u_{n\downarrow} \frac{\partial u^*_{n \uparrow}}{\partial x}+
u_{n\uparrow}\frac{\partial u^*_{n \downarrow}}{\partial x} \Bigr ]
-(1-f_n)
\Bigl[v_{n\uparrow} \frac{\partial v^*_{n \downarrow}}{\partial x}-
v_{n\downarrow} \frac{\partial v^*_{n \uparrow}}{\partial x}+
v^*_{n\uparrow} \frac{\partial v_{n \downarrow}}{\partial x}-
v^*_{n\downarrow} \frac{\partial v_{n \uparrow}}{\partial x} \Bigr ]\Biggr\rbrace, \\
&{ S}_z =\hspace{-.1cm}
-\frac{i}{2m}\sum_n \Biggl\lbrace f_n\Bigl[u_{n\uparrow}^* \frac{\partial u_{n \uparrow}}{\partial x}-
u_{n\uparrow} \frac{\partial u^*_{n \uparrow}}{\partial x}-
u^*_{n\downarrow} \frac{\partial u_{n \downarrow}}{\partial x}+
u_{n\downarrow}\frac{\partial u^*_{n \downarrow}}{\partial x} \Bigr ]
-(1-f_n)
\Bigl[-v_{n\uparrow} \frac{\partial v^*_{n \uparrow}}{\partial x}+
v^*_{n\uparrow} \frac{\partial v_{n \uparrow}}{\partial x}+
v_{n\downarrow} \frac{\partial v^*_{n \downarrow}}{\partial x}-
v^*_{n\downarrow} \frac{\partial v_{n \downarrow}}{\partial x} \Bigr ]\Biggr\rbrace.
\end{align}
\end{widetext}
In the case of $F$ layers with uniform magnetization,
there is no net spin current. The introduction of an
inhomogeneous magnetization texture however
results in a net spin current imbalance 
that is finite  even in 
the absence of a charge current.

\end{document}